\documentclass[reprint,superscriptaddress,amsmath,amssymb,pra,showpacs,floatfix]{revtex4-1}
\usepackage{graphicx}
\usepackage{dcolumn}
\usepackage{bm}
\usepackage{amsmath}
\usepackage{amssymb}
\usepackage{color}
\usepackage{hyperref}
\usepackage{times}

\hypersetup{  colorlinks=true,           linkcolor=blue,              citecolor=red,            urlcolor=blue  }

\begin{document}

\title{Nitrogen-Vacancy Center as Open-Quantum-System Simulator}
\author{Chao Lei}%
\thanks{These two authors contributed equally}
\affiliation{Hefei National Laboratory for Physics Sciences at Microscale and Department of Modern Physics, University of Science and Technology of China, Hefei, 230026, China}
\affiliation{Synergetic Innovation Center of Quantum Information and Quantum Physics, University of Science and Technology of China, Hefei, Anhui 230026, People's Republic of China}
\affiliation{Department of Physics, The University of Texas at Austin, Austin, Texas 78712,USA}
\author{Shijie Peng}%
\thanks{These two authors contributed equally}
\affiliation{Hefei National Laboratory for Physics Sciences at Microscale and Department of Modern Physics, University of Science and Technology of China, Hefei, 230026, China}
\author{Chenyong Ju}%
\affiliation{Hefei National Laboratory for Physics Sciences at Microscale and Department of Modern Physics, University of Science and Technology of China, Hefei, 230026, China}
\affiliation{Synergetic Innovation Center of Quantum Information and Quantum Physics, University of Science and Technology of China, Hefei, Anhui 230026, People's Republic of China}
\author{Man-Hong Yung}\altaffiliation{Corresponding author: yung@sustc.edu.cn}
\affiliation{Department of Physics, South University of Science and Technology of China, Shenzhen, Guangdong, 518055, People's Republic of China}
\author{Jiangfeng Du}\altaffiliation{Corresponding author: djf@ustc.edu.cn}
\affiliation{Hefei National Laboratory for Physics Sciences at Microscale and Department of Modern Physics, University of Science and Technology of China, Hefei, 230026, China}
\affiliation{Synergetic Innovation Center of Quantum Information and Quantum Physics, University of Science and Technology of China, Hefei, Anhui 230026, People's Republic of China}

\begin{abstract}
Quantum mechanical systems lose coherence through interactions with external environments---a process known as decoherence. Although decoherence is detrimental for most of the tasks in quantum information processing, a substantial degree of decoherence is crucial for boosting the efficiency of quantum processes, for example, in quantum biology. The key to the success in simulating those open quantum systems is therefore 
the ability of controlling decoherence, instead of eliminating it. Here we focus on the problem of simulating quantum open systems with Nitrogen-Vacancy centers, 
which has become an increasingly important platform for quantum information processing tasks. Essentially, we developed a new set of steering pulse sequences 
for controlling various coherence times of Nitrogen-Vacancy centers; our method is based on a hybrid approach that exploits ingredients in both digital and analog
 quantum simulations to dynamically couple or decouple the system with the physical environment. Our numerical simulations, based on experimentally-feasible 
parameters, indicate that decoherence of Nitrogen-Vacancy centers can be controlled externally to a very large extend.
\end{abstract}


\maketitle

%

{\bf Introduction---} A quantum simulator~\cite{Feynman1982,Lloyd1996,Buluta2009,Kassal2010,Yung2012c} is potentially a powerful tool for solving many-body problems that are not tractable 
by classical methods. Generally, there are two types of quantum simulators. The first type, called digital quantum simulator~\cite{Lloyd1996,Lanyon2011,Zhang2012c,Yung2010c,Li2011,Yung2014}, makes use of a general-purpose quantum computer, where quantum states are encoded with qubits 
and the dynamical evolution is programmed in a quantum circuit. The other kind of quantum simulators are called analog quantum simulators~\cite{Somaroo1999,Lamata2007,Gerritsma2010,Yung2016,Huh2016a}, where the Hamiltonian 
of the simulated quantum system is directly engineered in a dedicated quantum device, for example, trapped ions~\cite{Gerritsma2010,Shen2015} and optical lattices~\cite{Greiner2008}.

The main challenge of constructing a practical quantum simulator is to reduce the influence of environmental decoherence, a universal problem for all tasks in quantum information processing including quantum communication~\cite{Xu2013}. In practice, a quantum simulator is necessarily an {\it open quantum system}, where the underlying system-environment interaction~\cite{OpenQS2002} plays the main role in determining the performance of a quantum simulator. Furthermore, it is important to understand how a quantum simulator can simulate open quantum 
systems~\cite{OpenQS2002}, which are of fundamental importance for understanding many physical phenomena in, for example, quantum optics~\cite{Scully1997}, 
quantum measurement~\cite{Schlosshauer05}, and biological systems~\cite{Suess2014}.

In the literature, digital approaches of open-system quantum simulation~\cite{Bacon2001,Yung2010d,Xu2012,Sweke2014} have been 
theoretically studied and experimentally demonstrated. Similarly, analog quantum simulators of open quantum system has been theoretically 
proposed~\cite{Mostame2012} and experimentally investigated~\cite{Haeberlein2015}. An important approach to tackle the decoherence problem, called dynamical decoupling~\cite{Viola1999,Khodjasteh2005,Uhrig2007,Yang2008,Gordon2008,Uhrig2009,West2010}, 
have been developed to significantly eliminate the system-environment interactions~\cite{Breuer2016}, through a sequence of external pulses applied to the system. 
Experimental implementations of dynamical decoupling indicate that such approach is widely applicable to 
various experimental platforms~\cite{Biercuk2009,Du2009,Lange2010,Sagi2010,Averin2016,Bluhm2011,Medford2012,Ryan2010}. 
Moreover, an extension of dynamical decoupling is possible for universal 
quantum computation~\cite{West2010b,Souza2011} and other applications~\cite{Khodjasteh2010,Wolfowicz2014}.

Here we study the possibility of simulating {\it open} quantum systems through an extension of the idea of dynamical decoupling. More precisely, we developed a new set of  
decoupling pulse sequences that can control the coherence times of the off-diagonal matrix elements of the system density matrix. 

The pulse sequences of interest in this work are different from those in dynamical decoupling \cite{Viola1999,Khodjasteh2005,Uhrig2007}, which primary goal is to {\it decouple} the system from the influence of the environment. In other words, the goal of dynamical decoupling was to maintain the purity of the quantum system. Here we aims to {\it{control}} the decoherence by exploiting the existing environment, so that we can simulate the dynamics of an open quantum system without the need to maintain the purity of the system. Consequently, we can avoid the need of including extra ancilla qubits as in other digital approaches of simulating open quantum systems.

{\bf Full system controllability---} A NV center can be viewed as a 3-dimensional qudit with a Hamiltonian ${H_{NV}}$ of the following form \cite{Lange2010,Hanson2008,Hanson2006}:
\begin{equation}\label{NV}
    {H_{NV}} = DS_z^2 + \gamma {B_z}{S_z} \ ,
\end{equation}
where $ S_z $ is the 3-dimensional spin operator for the spin-1 particle $\{ m=1,0,-1\}$, $D$ is the zero-field splitting, $ \gamma = g{\mu _B} $ is the gyromagnetic ratio with ${\mu _B}$ 
the Bohr magnetism, $g$ the g-factor of electron, and ${B_z}$ is the static magnetic field applied along $z$ direction ([111] axis). Before we demonstrate how to control decoherence, we first show how to simulate a general $d$-dimensional system with a NV center. For a chosen basis $ \{ \left| m \right\rangle \}$, the Hamiltonian $H_S$ of a general $d$-dimensional system can be expressed as: ${H_S} = \sum\nolimits_{m = 0}^{d - 1} {{\varepsilon _m}\left| m \right\rangle \langle m|}  + \sum\nolimits_{m < n}^{d - 1} {({J_{mn}}\left| m \right\rangle \langle n| + J_{mn}\left| n \right\rangle \langle m|)}$,
where ${\varepsilon _m} = \left\langle m \right|{H_S}\left| m \right\rangle $ is the energy of the $m$-th state $\left| m \right\rangle$, and ${J_{mn}} = \left\langle m \right|{H_S}\left| n \right\rangle $ is the coupling between the $m$ and $n$-th state.  

For any given Hamiltonian $ \hat{H} $ and quantum state $ \left| \Psi \right\rangle $, and a unitary operator $ \hat{U} = e^{-i \hat{A} t} $ associated with a self-adjoint operator $ \hat{A} $, the corresponding quantum state in the rotating frame is given by, $ \left| {{\Psi_{\rm{rot}}}} \right\rangle  \equiv {\hat U^\dag }\left| \Psi  \right\rangle $.
We can obtain an effective Hamiltonian $ {H_{\rm{rot}}}$ in the rotational frame as follows: ${H_{\rm{rot}}} = {{\hat U}^\dag }\hat H\hat U - \hat A$.
In our case, we include two sets  of microwave pulses to a NV center. In the laboratory frame, the Hamiltonian in Eq.~(\ref{NV}) becomes: 
\begin{equation}\label{NVmicro}
    {H_{NV}^{mw}} = H_{NV} + \gamma {B_1}{S_x}\cos {\omega _1}t + \gamma {B_2}{S_x}\cos {\omega _2}t \ ,
\end{equation}
where $H_{NV}$ is the Hamiltonian shown as in Eq.~(\ref{NV}), $S_x$ is a $3 \times 3$ spin operator, $B_{1}$ and $B_{2}$ are the applied magnetic field along the $\hat{x}$ axis, and $\omega_{1}$ and $\omega_{2}$ are the frequencies of the applied microwave. 

The target Hamiltonian $H_S$ can be obtained by choosing a rotational frame reference where $\hat A = {\omega _1}\left| 1 \right\rangle \left\langle 1 \right| + {\omega _2}\left| { - 1} \right\rangle \left\langle { - 1} \right| $, 
which gives the following: ${\varepsilon _1} = D + \gamma {B_z} - {\omega _1}$, ${\varepsilon _2} = 0$, 
${\varepsilon _3} = D + \gamma {B_z} - {\omega _2}$, ${J_{12}} = \gamma {B_1}/2\sqrt 2 $, and ${J_{23}} = \gamma {B_2}/2\sqrt 2 $. When the coupling parameters are taken to be some real values, the $3\times3$ version of the Hamiltonian in Eq.~(\ref{NV}) contains 4 free parameters (apart from an overall shift of the total energy). 


{\bf Noises in Nitrogen-Vacancy (N-V) centers---} For a general open quantum system, the Hamiltonian $H$ can be divided into three parts: $H = H_{S} + H_{SB} + H_{B}$,
where $H_{S}$ is the Hamiltonian of quantum system, $H_{SB}$ is the Hamiltonian of the system-environment interaction, and $H_{B}$ is the Hamiltonian of 
environment (bath).
Normally, a NV center is subject to a local environment dominated by the surrounding nuclear spins  of $^{13}$C's and electron spins of P1 centers, which effectively produce a random magnetic field $b(t)$ to the NV center~\cite{Hanson2008,Lange2010}, i.e., ${H_{SB}} = b\left( t \right){S_z} = b\left( t \right)\left( {\left| 1 \right\rangle \left\langle 1 \right| - \left| -1 \right\rangle \left\langle -1 \right|} \right) $.

For the cases where the environment is dominated by the nuclear spins, the random fluctuation of the magnetic field can be regarded as stationary, i.e., $b(t)=b$, and is usually approximated as Markovian and Gaussian~\cite{Hanson2008}, i.e., with a probability distribution $\Pr (b) = {e^{ - {b^2}/2\sigma _b^2}}/\sqrt {2\pi } {\sigma _b}$,
where $ \sigma _b $ is the variance of the random magnetic field from the spin bath. For the cases where the noise come from electron spins instead, the random process of $b(t)$ can be approximated by the Ornstein-Uhlenbeck process~\cite{Lange2010}, with a correlation function $C(t)$ given by the following, $ C(t) = \left\langle {b(0)b(t)} \right\rangle  = {l^2}\exp ( - R\left| t \right|) $,
where $l$ describes the characteristic strength of the coupling of the NV center to the bath, and $R = 1/\tau_c$ is the transition rate, with $\tau_c$ being the correlation time of the spin bath~\cite{Lange2010}.

{\bf Strengthening decoherence---} We note that the evolution operator ${e^{ - iHt}}$ of the total system can be divided by many small time slices, $\Delta t \equiv t /n$, ${e^{ - iHt}} = \mathop {\lim }\nolimits_{n \to \infty } {\left( {{e^{ - i{H_S}\Delta t}} \ {e^{ - i{H_{SB}}\Delta t}} \ {e^{ - i{H_B}\Delta t}}} \right)^n}$. One way to strengthen decoherence, assisted by the environment, can be achieved as follows: first, turn off the system Hamiltonian momentarily (setting $H_S =0$), for a time period, $\lambda \Delta t$, where $\lambda >0$, i.e., ${e^{ - i({H_{SB}} + {H_B})\lambda \Delta t}}$. Then we allow the total system to evolve freely for a time period of $\Delta t$. The pattern is then repeated for $n$ times, i.e., ${( {{e^{ - i({H_S} + {H_{SB}} + {H_B})\Delta t}}{e^{ - i({H_{SB}} + {H_B})\lambda \Delta t}}} )^n}$.
Therefore, in the large-$n$ limit, we obtain an effective Hamiltonian as follows,
\begin{equation}\label{hameff}
    {H_{\rm eff}} = {H_S} + (1 + \lambda )({H_{SB}} + H_B) \ ,
\end{equation}
which contains an interaction term $H_{SB}$ amplified by a factor of $ (1 + \lambda) $. The side product is that the energies of the environment is also amplified. However, for the spin environments of NV centers, the effect can be ignored.
 
\begin{figure}[t]
  \includegraphics[width=\linewidth]{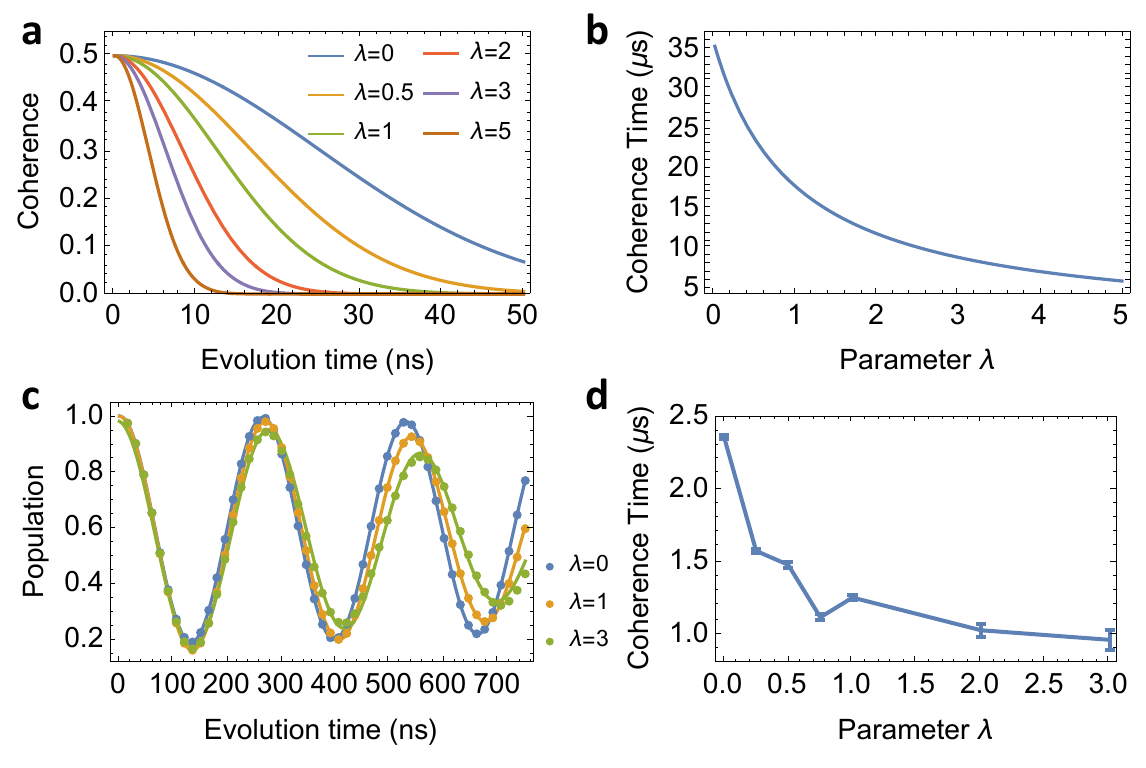}\\
  \caption{(Color online) (a) Coherence vs evolution time, the variance of the random magnetic field is set to be $0.2$ Gauss. (b) Coherence time vs $\lambda$. c)Evolution of population with time, the dots are simulated points and the curves are fitted one. 
(d) Coherence time got from the fitting curve in (c) vs $\lambda$ from 0 to 3, 
here the values of $\lambda$ in the calculation are 0, 0.25, 0.5, 0.75, 1, 2, 3. }\label{fig2lev}
\end{figure}

{\bf Controlling noises from nuclear and electron spin baths---} In the following we consider combining Trotter expansion with decoupling pulses to control the decoherence in NV centers. For NV centers in ultra-pure diamonds, the dominant decoherence source comes from the surrounding $^{13}$C spin bath~\cite{Hanson2008}, which is random but stationary within the time-scale of system dynamics. Consequently, for a qubit initialized in a pure state, $ \left| {\psi _0} \right\rangle = \alpha \left| 0 \right\rangle  + \beta \left| 1 \right\rangle  $, and $ H_S = 0$, 
the off-diagonal matrix element (or coherence), ${\rho _{12}} = \alpha {\beta ^*}\int {\Pr \left( b \right){e^{ - i2b(1 + \lambda )t}}db}$,
decays as follows: ${\rho _{12}} = \alpha {\beta ^*}{e^{ - 2\sigma _b^2{{(1 + \lambda )}^2}{t^2}}}$, which implies that the effective decoherence time $T_2$, can be controlled by the parameter $\lambda$,
\begin{equation}\label{t2s2lev}
{T_2}(\lambda ) = 1/\sqrt 2 {\sigma _b}(1 + \lambda ) .
\end{equation}
From Fig.\ref{fig2lev}(a), it is clear that tuning the parameter $\lambda$ can strengthen the decoherence, i.e., we can destroy the coherence via increasing the value of $\lambda$. The coherence time can be extract from above and shown in Fig.~\ref{fig2lev}(b).

In diamonds with nitrogen impurities, such as Type I, the electron spins also contribute to the decoherence of the system, which means that both nuclear and electron noise should be included. We investigate the decoherence of a two-level system numerically with experimental parameters of NV center. The simulated results are shown in Fig.\ref{fig2lev} (c) and (d), in which one microwave is applied. The parameters are as follow: The static magnetic field is $ B_z = 100 $ Gauss, 
the zero-field splitting is $ D = 2.87 GHz $, the frequency of applied microwave is a little away from the resonance frequency(which is 3.15 GHz) of NV center 
at the given static magnetic field with a detuning ${\rm{1}}{\rm{.9}} \times {\rm{1}}{{\rm{0}}^6}{\rm{/(1 + }}\lambda {\rm{)}}$ Hz and with a amplitude of 1.717 Gauss, 
$ \lambda $ is varied from zero (i.e. the $T_2^*$) to 3. the coherence time vs $\lambda$ is shown in  Fig.\ref{fig2lev} (d), it shows that the coherence
 time of NV center decrease when $ \lambda $ increases, which shows a effective control of decoherence caused by system-environment interaction.



\begin{figure}[t]
  \includegraphics[width=\linewidth]{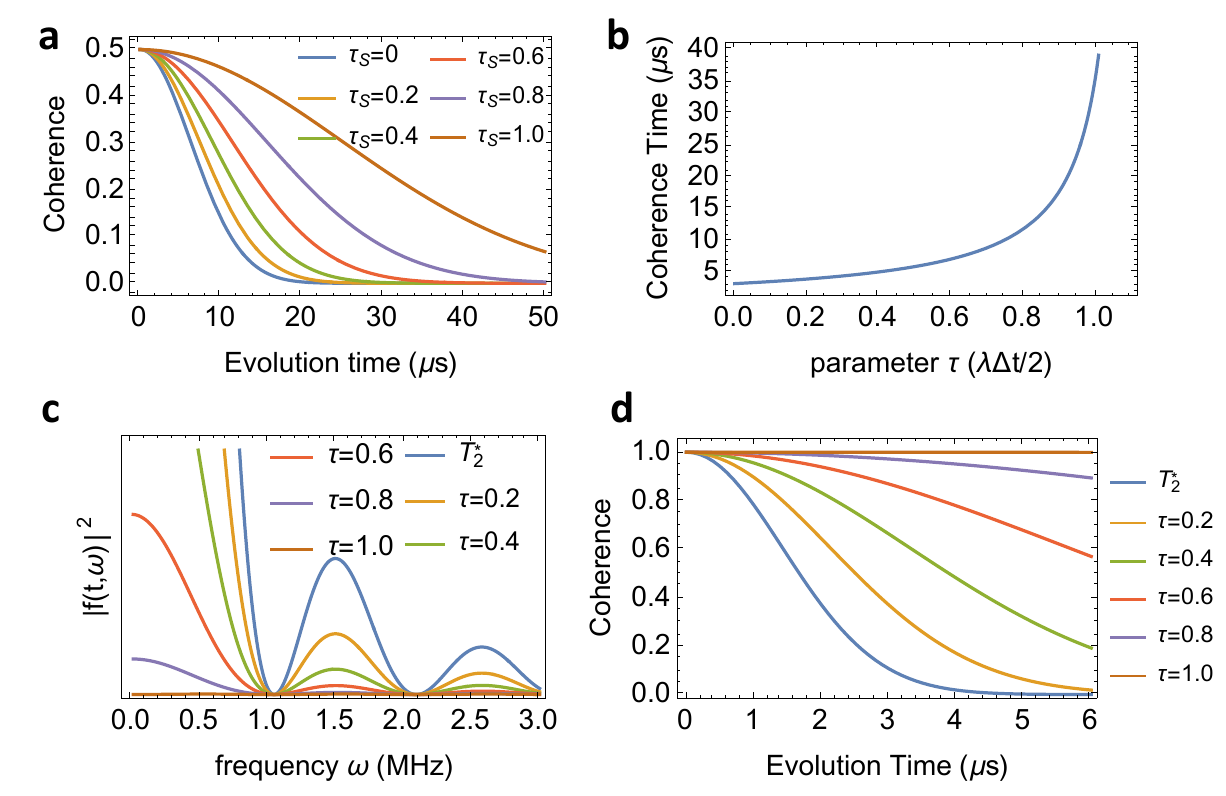}\\
  \caption{(Color online) (a) Coherence vs evolution time with different distance between two swap gate, the variance of the random magnetic field is set to be 0.2 $Gauss$ .
  (b) Coherence time vs $\tau$, which is defined in $ \mu \Delta t = \tau \lambda \Delta t/2 $.
  (c) is the values of $| {\tilde f(t,\omega )}|^2$ which represent the noise spectrum. 
  (d) The coherence vs time with different $\tau$. }\label{fig2levdyn}
\end{figure}

{\bf Weakening decoherence---} Let us consider a two-level system and a swap gate defined by: ${\rm{ }}{u_{12}} \equiv \sigma_x = \left( \begin{smallmatrix} 0&1\\ 1&0 \end{smallmatrix} \right)$. During the decoupling part of Trotter decomposition, we include the following evolution: ${\sigma _x}{e^{ - i{H_{SB}}{t_2}}}{\sigma _x}{e^{ - i{H_{SB}}{t_1}}} = {e^{ - i{H_{SB}}\left( {{t_1} - {t_2}} \right)}}$,
where we set $ t_1 = (\lambda - \mu) \Delta t $ and $ t_2 = \mu \Delta t $. The overall evolution becomes ${({e^{ - i({H_S} + {H_{SB}})\Delta t}}{e^{ - i{H_{SB}}(\lambda  - 2\mu )\Delta t}})^n}$, which implies the following effective Hamiltonian:
\begin{equation}\label{hameff2lev}
    {H_{\rm eff}} = {H_S} + (1 + \lambda  - 2 \mu ){H_{SB}}
\end{equation}
Suppose $ H_S = 0$, the coherence vs evolution time under Eq.~(\ref{hameff2lev}) becomes: $  {\rho _{12}} = \alpha {\beta ^*}{e^{ - 2\sigma _b^2{{( 1 + \lambda - 2 \mu)}^2}{t^2}}}$. This is shown in Fig.\ref{fig2levdyn} (a) and the coherence time is:
\begin{equation}\label{coh2levdyn}
T_2^{dd} = 1/\sqrt 2 {\sigma _b}(1 + \lambda  - 2\mu ) \ ,
\end{equation}
which is shown in Fig.~\ref{fig2levdyn}(b). Here we set the distance between two swap gate as $ \mu \Delta t = \tau \lambda \Delta t/2 $. When $\tau = 0$, it is equal to the case with no decoupling pulse, but when $ \tau = 1 $, the distance between the two swap gate is half of the $\lambda \Delta t$, 
which is the same as the CPMG pulse~\cite{Carr1954,Meiboom1958}. 

In the presence of electron spin noise, the coherence factor of a two-level system subject to dynamical decoupling is given~\cite{Cywinski2008} by, $  W(t) = | {\langle {\exp ( - i\int_0^t {b(t')f(t;t')dt'} )} \rangle } | = {e^{\chi (t)}}$, where $b(t)$ is the random noise, the function $ f(t;t') $ depends on the pulse sequence as, $ f(t;t') = \sum\nolimits_{k = 0}^n {{{( - 1)}^k}\theta ({t_{k + 1}} - t')} \theta (t' - {t_k})$,
with $\theta (t')$ the Heaviside step function, $t_0 = 0$ and $t_{n+1}=t$ is the total evolution time. 

Furthermore, the spectral density of the noise $C(\omega)$ is given~\cite{Lange2010} by, $ C(\omega) = l^2 \frac{2R}{R^2 + \omega^2} $,
which implies that $\chi (t) = \int_0^\infty  {\frac{{d\omega }}{{2\pi }}C(\omega ){{| {\tilde f(t,\omega )} |}^2}}$,
where $\tilde f(t,\omega ) = \int_{ - \infty }^\infty  {{e^{i\omega t}}f(t;t')dt}$. For our case, the square of Fourier transform of $ f(t;t') $ is found to be, $    {| {\tilde f(t,\omega )} |^2} = \frac{1}{{{\omega ^2}}}\frac{{1 - \cos \omega t}}{{1 - \cos \omega \delta }}
(6 + 2\cos \omega \delta  - 4\cos \omega {\delta _1} - 4\cos \omega {\delta _2})$,
where $\delta = \lambda \Delta t$,$\delta _1 = (\lambda - \mu ) \Delta t $ and $\delta _2 = \mu \Delta t$. The values of $| {\tilde f(t,\omega )} |^2$ are shown in Fig.~\ref{fig2levdyn}(c) with different values of $\tau$, which is defined in the following relation, $\mu \Delta t = \tau \lambda \Delta t /2$.
We obtained the coherence factor shown as in Fig.~\ref{fig2levdyn}(d), after applying a large cut-off frequency. It is obviously that when $\tau$ becomes larger, the coherence time becomes longer.

%


{\bf Fine-tuning decoherence for qudit---} For a general multi-level systems, i.e., qudit, we have an extra tool to fine-tuning the decoherence for different off-diagonal elements in the density matrix. Here we consider only the stationary noise from nuclear spin in three-level system. Let us consider applying the dynamical decoupling pulses, 
${u_{12}} = \left| 1 \right\rangle \left\langle 0 \right| + \left| 0 \right\rangle \left\langle 1 \right| + \left| -1 \right\rangle \left\langle -1 \right|$, are applied on only one channel, 
between $\left| m={ 1} \right\rangle $ and $\left| m={ 0} \right\rangle $, then the relevant part in evolution operator becomes: ${u_{12}} \ {e^{ - i{H_{SB}}{t_2}}} \ {u_{12}} \ {e^{ - i{H_{SB}}{t_1}}} \equiv {e^{ - i b(t) L({t_1},{t_2})}} $,
where $L({t_1},{t_2}) = \left| 1 \right\rangle \left\langle 1 \right|{t_1} + \left| 0 \right\rangle \left\langle 0 \right|{t_2} - \left| -1 \right\rangle \left\langle -1 \right|({t_1} + {t_2}) $.
Therefore, when we choose $t_1 < t_2$, we effectively make state $\left| 1 \right\rangle $ experience less dephasing then state $\left| 0 \right\rangle $, and vice versa.


As an example, we again set: $ t_1 = (\lambda - \mu) \Delta t $ and $ t_2 = \mu \Delta t $ , which gives $ {t_1} + {t_2} = \lambda \Delta t$. For any given initial state, $ {\psi _0} = \alpha \left| 1 \right\rangle  + \beta \left| 0 \right\rangle  + \gamma \left| { -1} \right\rangle $, if we set $ H_S = 0$, 
then the off diagonal elements of the associated density matrix decays as follows: ${\rho _{12}} =  \alpha {\beta ^*}{e^{ - \sigma _b^2{{(1 + \lambda  - 2\mu )}^2}{t^2}/2}} $, ${\rho _{13}}  =  \alpha {\gamma ^*}{e^{ - \sigma _b^2{{(2 + 2\lambda  - \mu )}^2}{t^2}/2}} $, and ${\rho _{23}} =  \beta {\gamma ^*}{e^{ - \sigma _b^2{{(1 + \lambda  + \mu )}^2}{t^2}/2}} $.
In other words, the coherence times of the off-diagonal elements are given by $T_2^{12} = \sqrt 2 /{\sigma _b}(1 + \lambda  - 2\mu )$, $T_2^{13} = \sqrt 2 /{\sigma _b}(2 + 2\lambda  - \mu )$, and $T_2^{23} = \sqrt 2 /{\sigma _b}(1 + \lambda  + \mu )$.

\begin{figure}[t!]
  \includegraphics[width=\linewidth]{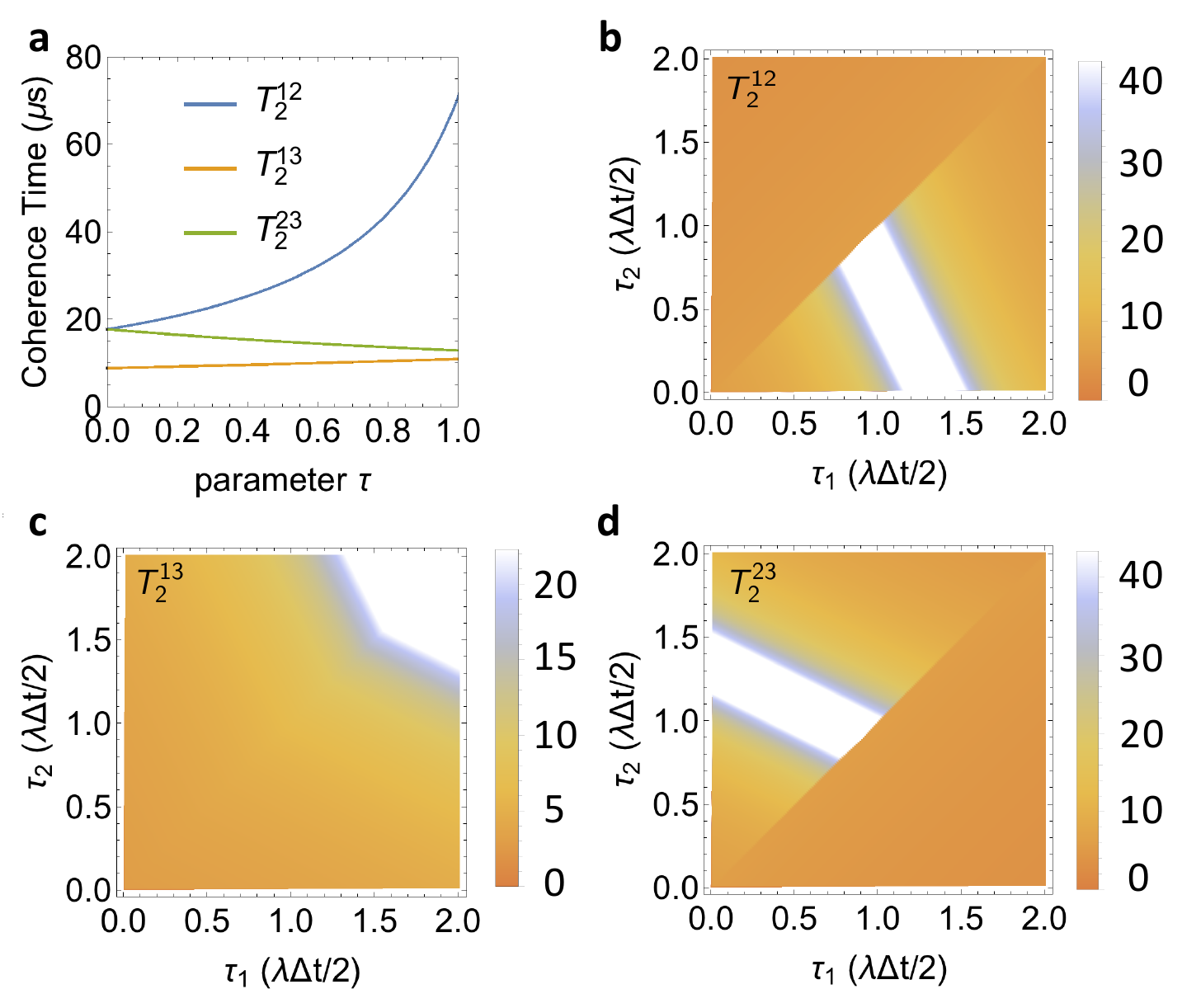}\\
  \caption{(Color online) 
(a)The coherence time vs $\tau$(which is defined in $ \mu \Delta t = \tau \lambda \Delta t/2 $), as $\tau$ increases, the coherence time $T_2^{12}$ increases 
and the other two coherence time remains almost the same, decrease or increase slightly.
(b)-(d) show the coherence time between the three levels, with 
(b) is the $T_2^{12}$, 
(c) the $T_2^{13}$ and 
(d)$T_2^{23}$, in which $\tau_1$ and $ \tau_2 $ are defined in $ \mu_{1(2)} \Delta t = \tau_{1(2)} \lambda \Delta t /2 $, 
the blue region means the coherence time is small and the red ones is the case the coherence time increases sharply. }\label{fig:5}
\end{figure}

The dependence of the coherence times with the parameter~$\mu$ is shown in Fig.~\ref{fig:5}(a). We see that the coherence time $T_2^{12}$ is more sensitive to the change of $\mu$, compared with the other coherence times.


Furthermore, additional dynamical decoupling pulses, e.g., ${u_{23}} = \left| 1 \right\rangle \left\langle 1 \right| + \left| -1 \right\rangle \left\langle 0 \right| + \left| 0 \right\rangle \left\langle -1 \right|$, 
can be applied, between $\left| m= { - 1} \right\rangle  \leftrightarrow \left| m = 0 \right\rangle $ and 
$ \left| m = 0 \right\rangle  \leftrightarrow \left| m = 1 \right\rangle  $, then we have (let $\mu_2 > \mu_1 $): $    {u_{23}} \ {u_{12}} \ {e^{ - i{H_{SB}}{t_3}}}{u_{12}}{e^{ - i{H_{SB}}{t_2}}}{u_{23}}{e^{ - i{H_{SB}}{t_1}}} \equiv {e^{ - i b(t) L({t_1},{t_2},{t_3})}}$,
where $ L({t_1},{t_2},{t_3}) = \left| 1 \right\rangle \left\langle 1 \right|(t_1 + t_2) - \left| 0 \right\rangle \left\langle 0 \right|(t_2 + t_3 ) - \left| -1 \right\rangle \left\langle -1 \right|({t_1} - {t_3}) $.
Suppose we set $ t_1 = (\lambda - \mu_2) \Delta t $, $ t_2 = (\mu_2 - \mu_1) \Delta t $ and $ t_3 = \mu_1 \Delta t $ (shown in Fig.~\ref{fig:5}a) with $0 \le \mu_1 \le \lambda $ 
and $ \mu_1 \le \mu_2 \le \lambda $. It gives $ t_1 + t_2 + t_3 = \lambda \Delta t $. Consequently, for any initial state,
$ {\psi _0} = \alpha \left| 1 \right\rangle  + \beta \left| 0 \right\rangle  + \gamma \left| { -1} \right\rangle $, the time-dependent off-diagonal elements of the density matrix are given by: $   {{\rho _{12}} (t)  = \alpha {\beta ^*}{e^{ - \frac{1}{2}\sigma _b^2{{(1 + \lambda  - (\mu_1 - \mu_2))}^2}{t^2}}}} $, and similarly for $  {{\rho _{13}} (t)  = \alpha \, {\gamma ^*} \, {e^{ - \frac{1}{2}\sigma _b^2{{(2 + 2\lambda  - (2\mu_1 + \mu_2))}^2}{t^2}}}}$ and $   {{\rho _{23}} (t)  = \beta {\gamma ^*}{e^{ - \frac{1}{2}\sigma _b^2{{(1 + \lambda  - (\mu_1 + 2\mu_2))}^2}{t^2}}}} $,
which means that the coherence times of the off-diagonal elements are given by, $ {T_2^{12} = \sqrt 2 /{{{\sigma _b}(1 + \lambda  - (\mu_1 - \mu_2))}}} $ for $\rho_{12} (t)$, 
$ {T_2^{13} = \sqrt 2 /{{{\sigma _b}(2 + 2\lambda  - (2\mu_1 + \mu_2))}}} $ for $\rho_{13}(t)$, and 
$ {T_2^{23} =  \sqrt 2 /{{{\sigma _b}(1 + \lambda  - (\mu_1 + 2\mu_2))}}} $ for $\rho_{23}(t)$.

The coherence times with parameters $ \mu_{1(2)} $ are shown in Fig.\ref{fig:5}(b)-(d)(in which the case of $\tau_2 \le \tau_1 \le 2  $ is also included, 
where $\tau_1$ and $ \tau_2 $ are defined in $ \mu_{1(2)} \Delta t = \tau_{1(2)} \lambda \Delta t /2 $). In the plots, the blue region means the coherence time is small and the red ones is the case the coherence time increases sharply. It shows that the coherence between different levels can be tuned though 
changing the insert time of the two decoupling time, which is similar with the case of only one channel is applied with decoupling pulse. The decoupling pulses are applied on both channels; the coherence time $T_2^{13}$ increases when $\mu_1$ and $\mu_2$ are close to $\lambda $ (shown in Fig.\ref{fig:5}(c)), while the other two coherence time remains essentially the same as the one before applying the decoupling pulses.

Finally, further generalization of our method to $d$-dimensional ($d \geq 4$) systems is possible. Following the previous results, the decoherence of different off-diagonal elements can be controlled by the following sequence: $\prod\nolimits_{i < j} {{u_{ij}}} \, \prod\nolimits_{i < j} {[ \ {e^{ - i{H_{SB}}{t_{ij}}}} \ {u_{ij}} \ ]} \, {e^{ - i{H_{SB}}{t_0}}} \, {e^{ - i({H_S} + {H_{SB}})\Delta t}}$,
where $t_{ij}$'s are the adjustable waiting time before the swap gate between $i$ and $j$ level is applied after the next swap gate $    {u_{ij}} = I + \left| i \right\rangle \langle j| + \left| j \right\rangle \langle i| - \left| i \right\rangle \langle i| - \left| j \right\rangle \langle j| $,
where $t_0$ is included as the waiting time before the first swap gate applied.

{\bf Conclusion---} In conclusions, in this work, we have presented a new method that can engineer the environment induced decoherence by combining the Trotter decomposition and decoupling 
pulses. The scheme exploits the intrinsic decoherence from the environment, and contains the benefits of the university of digital quantum simulation and also the efficiency of analog quantum simulation. This hybrid simulation method is numerically tested for NV centers with two and three energy levels. Our results indicate that such a scheme is experimentally feasible.

{\bf Acknowledgements}
This work was supported by the National Key Basic Research Program of China (Grant No. 2013CB921800), the National Natural Science Foundation of China (Grant Nos. 11227901 and 11405093), 
and the Strategic Priority Research Program (B) of the CAS (Grant No. XDB01030400).

\bibliography{HybridQSref}

\end{document}